\documentclass[aps,prl,twocolumn,groupedaddress]{revtex4}
\usepackage{epsfig,enumerate,amsmath,amstext,amsthm,amssymb}
\usepackage{graphicx}

\def\a{{\alpha}}

\def\b{{\beta}}

\def\lam{{\lambda}}

\def\xx{\mathbf{x}}
\def\yy{\mathbf{y}}

\newcommand{\footnoteremember}[2]{\footnote{#2}\newcounter{#1}\setcounter{#1}{\value{footnote}}}
\newcommand{\footnoterecall}[1]{\footnotemark[\value{#1}]}

\begin{document}

\title{Divide and concur: A general approach to constraint satisfaction}
\author{Simon Gravel}
\affiliation{Laboratory of Atomic and Solid-State Physics, Cornell
University, Ithaca, NY, 14850-2501 USA}
\author{Veit Elser}
\affiliation{Laboratory of Atomic and Solid-State Physics, Cornell
University, Ithaca, NY, 14850-2501 USA}

\date{\today}

\begin{abstract}
Many difficult computational problems involve the simultaneous satisfaction of multiple constraints which are individually easy to satisfy. Such problems occur in diffractive imaging, protein folding, constrained optimization (e.g., spin glasses), and satisfiability testing. We present a simple geometric framework to express and solve such problems and apply it to two benchmarks. In the first application (3SAT, a boolean satisfaction problem), the resulting method exhibits similar performance scaling as a leading context-specific algorithm (\textsc{walksat}). In the second application (sphere packing), the method allowed us to find improved solutions to some old and well-studied optimization problems. Based upon its simplicity and observed efficiency, we argue that this framework provides a competitive alternative to stochastic methods such as simulated annealing.
\end{abstract}

\pacs{05.10.-a, 02.70.-c, 05.45.Gg}

\maketitle

Difficult problems can often be broken down into a collection of smaller, more tractable, subproblems. This is the basis of the \textit{divide and conquer} approach, which applies when the initial problem and the subproblems have a similar structure, and the global solution can be retrieved from the solutions to the subproblems. Divide and conquer as a rule leads to very efficient algorithms. However, many difficult problems do not fit such an efficient framework.

For example, consider the problem of determining the three-dimensional structure of a complex molecule given clues about the distances between particular pairs of atoms (from knowledge of chemical bonds, NMR measurement, etc.).
As subproblems we might consider the substructures formed by small groups of atoms, since finding substructures satisfying local constraints is usually not challenging. However, the location and orientation in space of the substructures depends intricately and sensitively on their collective arrangement.
Because the division into subproblems in this case does not lead to a practical algorithm, molecular geometry problems are usually transformed into optimization problems through the definition of a global cost function, and are then solved through stochastic optimization methods such as simulated annealing.

In this Letter we introduce a general method for solving constraint problems that takes advantage of the division into subproblems. In broad terms the method differs from stochastic searches in that the configurations explored are generated iteratively and \textit{deterministically}. Each iterative step is defined by two fundamental operations. In the first operation, the problem is divided into its constituent constraints, which are then solved independently, ignoring possible conflicts between different constraints. In the second operation, conflicts between constraints are resolved regardless of the satisfaction of the constraints. By a judicious application of these two operations, 
we obtain a search strategy which, at each step, solves all the subproblems separately and at the same time seeks to resolve conflicts between their solutions. We call this method \textit{divide and concur} (D\&C).

The D\&C approach can be applied to a wide range of problems, both discrete and continuous. We first show how D\&C is applied to the boolean satisfiability problem (SAT), a standard benchmark in computer science. In this problem, the D\&C approach exhibits similar scaling behavior to \textsc{walksat}, a leading SAT solver which outperforms general-purpose algorithms such as simulated annealing \cite{selman1996lss}. As a second example we study continuous sphere packing problems, which are formally similar to the molecular geometry example mentioned above. The D\&C approach matched or improved upon the best known packings in some well-studied, two-dimensional problems. In 10 dimensions it also discovered an interesting new sphere arrangement related to quasicrystals. The D\&C approach therefore combines the advantages of general purpose algorithms (versatility, simplicity) with the performance of special purpose algorithms (such as \textsc{walksat}).

In D\&C the individual constraints are first expressed as subsets of a Euclidean space $K$, thereby transforming the constraint satisfaction problem into the geometrical problem of finding a point in the intersection of multiple sets. The Euclidean space provides the setting to define distance-minimizing \textit{projections} to each of the $N$ constraint sets. The projection operators $\{P_i\}_{i=1\ldots N}$ will be the building blocks of the algorithm. Starting from an initial guess, one uses the projections to probe the constraint sets and update the guess. This idea has been studied extensively in the context of \emph{convex} constraint sets \cite {bauschke2004fba}.

Given $N$ \emph{primary} constraints expressed as subsets of $K,$ we first define the product space $K^{N},$ consisting of $N$ copies (or \emph{replicas}) of $K$ \footnoteremember{space}{In problems where each projection $P_i$ acts nontrivially only on a limited subset $K_i$ of $K$, the product search space may be reduced to $K_1\otimes K_2\otimes \cdots \otimes K_{N}$, with obvious performance gains. If a variable is involved in $N_i$ constraints, there will be $N_i$ copies, or \emph{variable replicas}, of this variable in the reduced search space.} . We then define, \emph{in the product space}, the `divide' constraint $D$ (enforcing one primary constraint on each replica) and the `concur' constraint $C$ (enforcing replica concurrence) \cite{Pierra,bauschke2004fba}.
%Starting with a set of $N$ \emph{primary} constraints and their associated projections $\{P_i\}_{i=1\ldots N}$, each acting in the Euclidean space $K$, we define the product space $K^{N}$ consisting of $N$ copies (or \emph{replicas}) of $K.$ We then define, in the product space, the `divide' constraint $D$ (enforcing the decoupled primary constraints) and the `concur' constraint $C$ (enforcing replica concurrence).
The associated projections, acting on $\mathbf{y}=\xx^{(1)}\otimes\xx^{(2)}\otimes\cdots\otimes\xx^{(N)}$, are
\begin{equation}\label{direct}
P_D(\mathbf{y})=P_1(\xx^{(1)})\otimes P_2(\xx^{(2)})\otimes \cdots \otimes P_{N}(\xx^{(N)}),
\end{equation}
which acts separately on each of the replicas, and
\begin{equation} \label{average}
P_C(\mathbf{y})=\bar \xx\otimes \bar \xx\otimes \cdots \otimes \bar \xx,
\end{equation}
which replaces the value of each replica by the average value $\bar\xx$ of all the replicas. In defining the concurrence projection, different weights $\lam_i$ may be assigned to different constraints, i.e.,  $\bar \xx=\sum_i(\lam_i \xx^{(i)})/\sum_i \lam_i$ \footnote {In the reduced search space, the average should be taken only on the variables in the reduced space. If $V_k$ is the index list of the constraints in which variable $k$ appears,  $\bar x_k=\sum_{i\in V_k}(\lam_i x_k^{(i)})/\sum_{i\in V_k} \lam_i$.}.
Changing the weights is equivalent to changing the metric of the product space; this possibility will prove beneficial even in problems where all the constraints are formally equivalent. Note that both projections, $P_C$ and $P_D$, act in a highly parallel sense: either by treating independently each replica ($P_D$), or by treating independently  each variable across the replicas ($P_C$).

Through the product space construction the original constraint problem has been expressed as the problem of finding a point in the intersection of two sets, both of which have easily implemented projection operators. To proceed, we need a search strategy that can use a pair of projection operators ($P_a$ and $P_b$) to seek the intersection of two sets. The simplest approach is the alternating projections scheme, where $\mathbf{y}_{n+1} = P_a(P_b(\mathbf{y}_n))$ \cite{Kaczmarz1937,bauschke2004fba}.  Despite its success with convex constraints and some nonconvex problems, the alternating projections scheme is prone to getting stuck at fixed points which do not correspond to solutions.

The difference map (DM) is an improvement upon alternating projections which emerged in response to the nonconvex constraints arising in diffractive imaging (the phase problem)\cite{elser2003pri,ThesePT}. It is defined by a slightly more elaborate set of rules, namely
\begin{equation}\label{DMgeneral}
\begin{split}
\yy_{n+1}=&\yy_n+\b \left(P_a\circ f_b(\yy_n)-P_b\circ f_a(\yy_n)\right)\\
f_i(\yy_n)=&(1+\gamma_i) P_i(\yy_n)-\gamma_i \yy_n~~~~~~~i=a,b,
\end{split}
\end{equation}
with $\gamma_a=-1/\beta$ and $\gamma_b=1/\beta.$ The parameter $\beta$ can have either sign and is chosen to improve performance. If the iteration reaches a fixed point $\yy^*$, an intersection of the constraint sets has been found. The solution $\yy_\mathrm{sol}$ is obtained from the fixed point using $\yy_\mathrm{sol}=P_a\circ f_b(\yy^*)=P_b\circ f_a(\yy^*)$. Note that the fixed point itself is not necessarily a solution; in fact, there typically is a continuum of fixed points associated with every solution.

The DM was recently used to solve a variety of difficult computational problems, including protein folding \cite{ivan1, ivan2}, boolean satisfiability, Diophantine equations, graph coloring, and spin glasses \cite{SSelser2007}. Most of these applications relied on a decoupling of constraints through the use of a dual set of variables, as in linear programming. The \textit{divide and concur} approach we introduce here, defined by the use of the difference map with $P_a=P_C$ and $P_b=P_D,$ is significantly more versatile and systematic.

The boolean satisfiability problem 3SAT is one of the most extensively studied problems in constraint satisfaction. The challenge is to find an assignement for $N_v$ boolean variables that satisfies a list of $N_c$ boolean constraints, or \emph{clauses}. Each clause is an OR statement involving three \emph{literals}, $\ell_{1}\vee \ell_{2} \vee \ell_{3},$ where each literal $\ell_{i}$ represents either one of the $N_v$ boolean variables or its negation.  

A D\&C formulation of 3SAT is obtained by associating a real-valued search variable to each 3SAT literal, where the values $\{1,-1\}$ are taken to mean $\{\mbox{True},\mbox{False}\}$. The constraint $D$ requires that each clause is satisfied; that is, each literal must have value $\pm1$, with at least one literal per clause having value $1$. In other words, to each clause corresponds a variable triplet, which is projected by $P_D$ to the nearest of the seven satisfying assignments for this clause. Geometrically, these correspond to seven vertices of a cube. In this application $P_C$ ensures that all literals associated with the same boolean variable concur (with due regard to negations). Since each constraint (clause) involves only three variables, the reduced search space\footnoterecall{space} has dimension $3 N_c$. For simplicity, we give equal weight to each constraint ($\lambda_i=1$).

\begin{figure}
 
%{\includegraphics[width=\columnwidth]{linregress.eps}}
\includegraphics[width=\columnwidth]{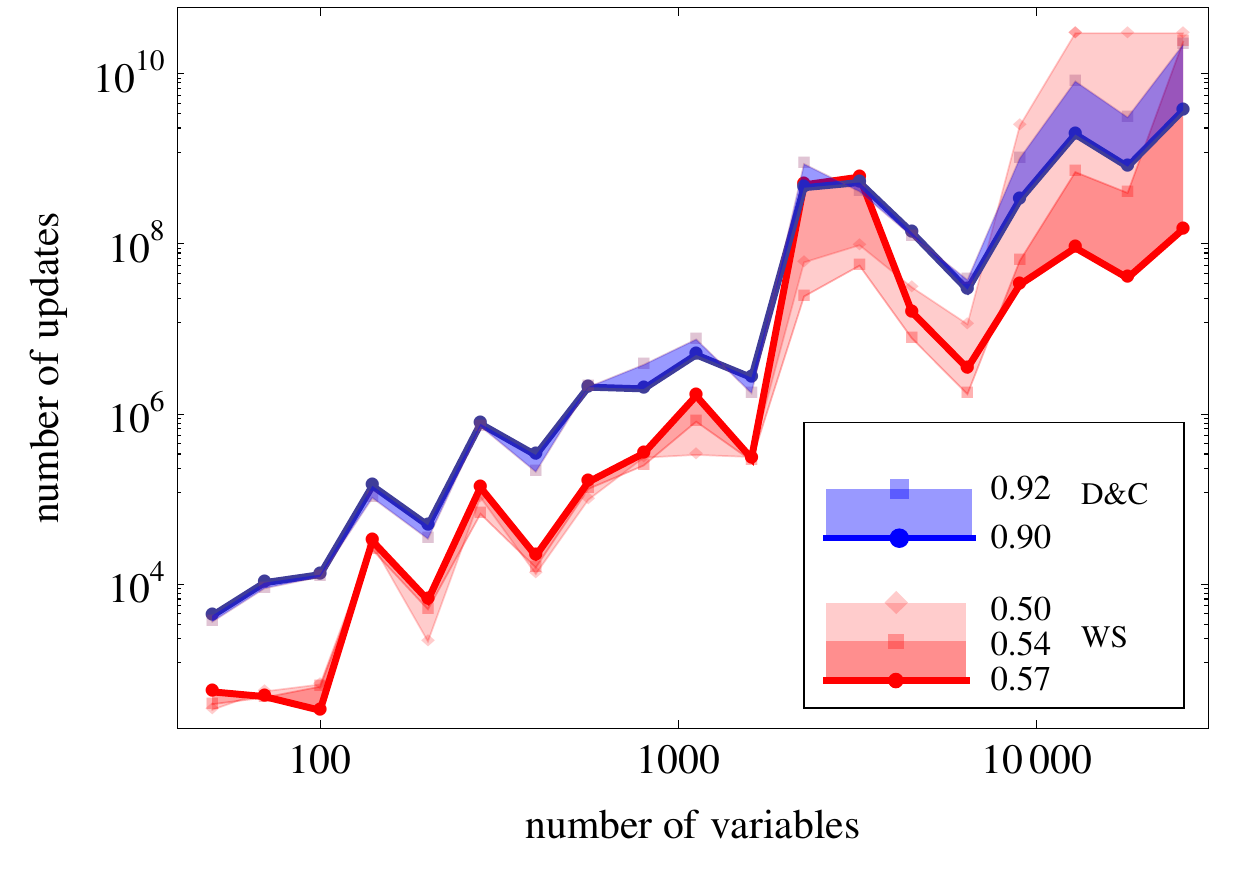}
%\scalebox{.65}{\includegraphics{shade2.pdf}}

\caption{Median number of variable updates needed to find a solution for \textsc{walksat} (WS) and divide and concur (D\&C) on the \emph{same} set of random 3SAT instances with $\alpha=4.2$. Each median was calculated by solving the same instance 10 times starting from different random initial guesses, for parameter values $\beta=0.9$ (D\&C) and $p=0.57$ (\textsc{walksat}). Variations resulting from changing $\beta$ and $p$ are indicated by the shaded areas; both methods exhibit parameter sensitivity for problems with more than $10^4$ variables. A point at the top edge indicates that the median exceeded the cutoff on the number of updates, $3\times 10^{10}.$  \label{regress}}
\end{figure}

We compared the performance of the D\&C algorithm with \textsc{walksat} \cite{selman1996lss} on a collection of 3SAT problem instances ranging from $N_v=50$ to $N_v=25600$, with fixed ratio $\alpha\equiv N_c/N_v= 4.2$, a value for which randomly generated instances are expected to be difficult \cite{cheeseman1991rhp}. Random instances were generated using the program \textsc{makewff} (distributed with \textsc{walksat}), and instances that were not solved by either \textsc{walksat} or the D\&C algorithm were discarded. Each algorithm was applied 10 times to each instance, starting from different random initial conditions. The median number of variable updates required to find the solution is plotted in Figure \ref{regress}. The number of variable updates in \textsc{walksat} equals the total number of flips of the boolean variables. In the D\&C algorithm it is the total number of nonzero updates of any of the real-valued search variables (literals). 

Fig. \ref{regress} shows that \textsc{Walksat} (with the `noise' parameter fixed at the value $p=0.57$) and the D\&C algorithm (with $\beta=0.9$) have similar performance behavior. Not only do they both find the same problems easy and the same problems hard (which is not unexpected), but the scaling of the number of variable updates needed to reach the solution, as a function of problem size, is also similar. Such a similarity is surprising, considering the difference in search strategies. 

\textsc{walksat} uses pseudorandom processes (or `noise') to update the variables asynchronously. In D\&C, on the other hand, the update rule is completely deterministic and is applied synchronously to many variables.
Fig. 1 also shows that choosing suboptimal parameters for either algorithm
results in rapid performance degradation for large problem sizes. 
Even though the scaling of the variable updates are similar for \textsc{walksat} and D\&C, our implementation of D\&C required significantly more CPU time (between $4$ and $200$ times, depending on the instance) than \textsc{walksat}. Work on an optimized implementation of the D\&C algorithm is in progress and should allow an easier exploration of the behavior of the method for larger problem sizes.

%\showthe\columnwidth

%\section{A continuous example: sphere packing}

%The problem of packing $n$ hard spheres in a finite $D$-dimensional volume is in many ways ideally suited for illustrating the present method. 

Another constraint problem which has been extensively studied is the packing of $n$ spheres in a finite $D$-dimensional volume (see, e.g., Refs. \cite{conway1999,stephenson2005icp,szabo2007} and references therein). The constraint formulation of this problem is more directly
geometrical than boolean satisfiability. Since each sphere must avoid $n-1$ other spheres and lie within a certain volume, there are altogether $n$ constraints per sphere. %Simultaneous satisfaction of these constraints cleary yields a solution to the packing problem. 
The reduced search space\footnoterecall{space} requires one $D$-dimensional variable replica for every sphere participating in a constraint, for a net search space dimensionality of $Dn^2.$ 

Within the framework of D\&C there is a formal similarity in the constraint structure of packing spheres and 3SAT. Just as every boolean variable is constrained by each of the clauses where it occurs, every sphere in a packing has a volume exclusion relationship with each of the other spheres in the packing: $\|\xx_a-\xx_b\|> m_{ab}$.
This similarity and the success of D\&C with 3SAT is strong motivation to apply D\&C to the sphere packing problem.

% (since each of the $n,$ $D$-dimensional spheres occurs in $n$ different constraints). 

Near the solution of any $n$-sphere packing problem, the number of relevant exclusion constraints (contacting pairs) grows only as $n$ (for fixed $D$) while the total number of constraints is $O(n^2)$. In the D\&C approach it is possible to increase the weight of these relevant pairs by dynamically adjusting the corresponding metric weight $\lam_{ab}$. This results in considerable performance improvement. At the end of each DM step we used $\lam_{ab}\rightarrow \sigma \lam_{ab}+ (1-\sigma)\exp\left(-\a\, d_{ab}\right)$, 
where $d_{ab}$ is the current distance between the pair\footnoteremember{pair}{To get a unique pair distance one uses coordinates given by the concurrence term of the difference map, $P_C\circ f_D$.}. We used the value $\sigma=0.99$ to ensure that the metric update is quasi-adiabatic (i.e., slow on the time scale of variable updates), and $\a \simeq 30$.

%\subsection{results}

We first consider packings of $n$ equal disks of diameter $m$ in a unit square, and take as a starting point the best known packing diameters $m^*$ from Ref. \cite{szabo2007}. This problem is quite challenging, due to the coexistence of many different arrangements with similar density. We tested the D\&C algorithm for each value of $n$ in the range 2-200. For each $n,$ we generated up to $400$ random initial guesses. For each initial guess, a small value of the diameter $m$ was chosen, and a packing was sought. When a solution was found, $m$ was increased, and the process was repeated until the algorithm failed to find a packing%after a predefined termination criterion
, or until the best known packing diameter $m^\star$ was reached. In the latter case the target was increased beyond $m^\star$ with the hope of finding a denser packing. No information about the known packings was used, apart from their densities.

\begin{figure}

\scalebox{.75}{\includegraphics{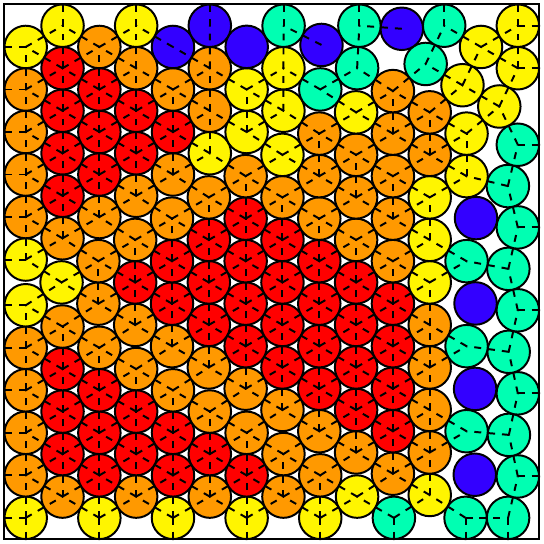}}
\scalebox{.75}{\includegraphics{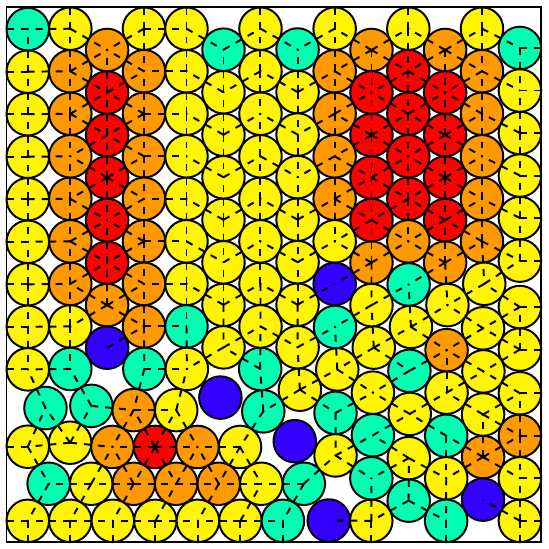}}
\caption{An example of an improved packing for 169 disks in a square found by the D\&C algorithm. The figure on the left shows the previously best known packing \cite{szabo2007}, with density $0.8393.$ The density of the improved packing shown on the right is $0.8399$. Contacts are shown with dotted lines; colors indicate the number of contacts. \label{pack}}
\end{figure}

For $143$ of the $197$ values of $n$ a packing with diameter close to the optimal packing ($m>m^\star-10^{-9}$) was found. More surprisingly, improved packings were found in $38$ cases. The smallest $n$ for which an improved packing was found is $91$. The largest improvement was for $n=182,$ for which a packing was found with $m=m^\star+4.6\times10^{-5}.$ For 28 values of $n$ a packing was found with $m>m^\star+1\times10^{-6}.$ An example of such an improved packing is shown in Figure \ref{pack}. % See EPAPS Document No. XXX for figures and coordinates of the improved packings.  

%Our purpose was not to perform a thorough investigation of dense sphere packings, or to carefully optimize the searching procedure.
% We conjecture that many additional improved packings can be obtained by using the methodology described above from additional initial guesses. 

%\subsection{Application too the kissing number problem}

When packing many disks the optimization challenge is easy to identify as a contest
between close-packing in the bulk and an efficient match to the boundary. In higher
dimensions the structure of the solution is not so easily characterized, and we can
look to the D\&C method as an unbiased tool for exploration. A classic problem in
geometry is to determine \textit{kissing numbers} $\tau_D$: the maximum number of
unit spheres that can be packed in $D$-dimensions, so that each contacts a given
unit sphere. Early investigations of this problem were stimulated by a debate
between Newton and Gregory, who disputed the value of $\tau_3$. The only known
kissing numbers are $\tau_1=2$, $\tau_2=6$, $\tau_3=12$, $\tau_4=24$, $\tau_8=240$,
and $\tau_{24}=196560$. In dimension 1-8, and also 16-24, the best known lower bounds
on $\tau_D$ are given by the number of minimal vectors in the unique laminated lattice
of the same dimension\cite{conway1999}. For dimension 9-15 the best bounds are obtained
from constructions based on error-correcting codes\cite{conway1999}. Discoveries of novel
packings in higher dimensions has for the most part been achieved through
mathematical inspiration. Unbiased searches, defined only by the basic constraints,
have to our knowledge not been attempted beyond dimension 5\cite{nurmela1995csc}. This raises the possibility that interesting packings in high
dimensions may have escaped detection only for lack of imagination.

With minimal adjustment to the above procedure for finding disk packings, we were
able to find kissing arrangements as good as the best known in dimension 2-4, 6, and
8. After introducing just the assumption of inversion symmetry, optimal packings
were obtained in all dimensions up to 8. Our searches in higher dimensions have so
far revealed an interesting new packing in dimension 10. It is easy to understand
why this packing was missed. Constructions based on integral lattices and
error-correcting codes all have the property that the cosine of the angle subtended
by any two spheres is rational. The packing of 378 spheres discovered by the D\&C
algorithm has all cosines in a set that includes irrational numbers: $\{\pm 1, \pm 1/2,
(\pm 3\pm\sqrt{3})/12, 0\}$. An analysis of the coordinates obtained by the
algorithm has revealed that these 378 sphere positions are expressible as unique integer multiples of a basis of 12 vectors. The construction has a strong relationship to quasicrystals, where the excess dimension of the basis accounts for irrational relationships in the geometry. The algorithm, of course, had no knowledge of
quasicrystal geometry.

This `irrational' structure emerged as soon as the number of spheres was increased above 372, the largest known kissing number for 10-dimensional lattices \cite{conway1999}. The same irrational arrangement was also found for up to 384 spheres; the 6 additional spheres were accomodated in holes of the structure (and have continuously variable cosines). %Coordinates for this arrangement can be found in EPAPS Document No. XXX.
Finally, the algorithm has so far been unsuccessful in discovering the best known kissing arrangement in 10 dimensions, with kissing number 500.

The divide and concur approach provides a natural framework in which to address various hard computational problems. In two benchmark applications, 3SAT and  sphere packing, the D\&C approach compares with, and in some cases improves upon, state-of-the-art specialized methods. The uniform mechanism provided by the difference map, for finding solutions given a pair of constraint projections, makes the D\&C approach almost as easy to implement as general-purpose sampling algorithms such as simulated annealing. Most of the problem-specific development needed, in this framework, is the definition of the appropriate projection operators. We believe the latter are able to exploit important elements of the problem structure not accessed by stochastic sampling, and that this accounts for the superior performance of D\&C.

We acknowledge useful discussions with Y. Kallus, D. Loh, I. Rankenburg, and P. Thibault. This work was supported by grant NSF-DMR-0426568.

%\bibliographystyle{apsrev}
%\bibliography{../references/replicas}

%\bibliography{replicas}

\begin{thebibliography}{14}
\expandafter\ifx\csname natexlab\endcsname\relax\def\natexlab#1{#1}\fi
\expandafter\ifx\csname bibnamefont\endcsname\relax
  \def\bibnamefont#1{#1}\fi
\expandafter\ifx\csname bibfnamefont\endcsname\relax
  \def\bibfnamefont#1{#1}\fi
\expandafter\ifx\csname citenamefont\endcsname\relax
  \def\citenamefont#1{#1}\fi
\expandafter\ifx\csname url\endcsname\relax
  \def\url#1{\texttt{#1}}\fi
\expandafter\ifx\csname urlprefix\endcsname\relax\def\urlprefix{URL }\fi
\providecommand{\bibinfo}[2]{#2}
\providecommand{\eprint}[2][]{\url{#2}}

\bibitem[{\citenamefont{Selman et~al.}(1996)\citenamefont{Selman, Kautz, and
  Cohen}}]{selman1996lss}
\bibinfo{author}{\bibfnamefont{B.}~\bibnamefont{Selman}},
  \bibinfo{author}{\bibfnamefont{H.}~\bibnamefont{Kautz}}, \bibnamefont{and}
  \bibinfo{author}{\bibfnamefont{B.}~\bibnamefont{Cohen}},
  \bibinfo{journal}{DIMACS Series in Discrete Mathematics and Theoretical
  Computer Science} \textbf{\bibinfo{volume}{26}}, \bibinfo{pages}{521}
  (\bibinfo{year}{1996}).

\bibitem[{\citenamefont{Bauschke et~al.}(2004)\citenamefont{Bauschke,
  Combettes, and Luke}}]{bauschke2004fba}
\bibinfo{author}{\bibfnamefont{H.}~\bibnamefont{Bauschke}},
  \bibinfo{author}{\bibfnamefont{P.}~\bibnamefont{Combettes}},
  \bibnamefont{and} \bibinfo{author}{\bibfnamefont{D.}~\bibnamefont{Luke}},
  \bibinfo{journal}{J. Approx. Theory} \textbf{\bibinfo{volume}{127}},
  \bibinfo{pages}{178} (\bibinfo{year}{2004}).

\bibitem[{\citenamefont{Pierra}(1976)}]{Pierra}
\bibinfo{author}{\bibfnamefont{G.}~\bibnamefont{Pierra}}, in
  \emph{\bibinfo{booktitle}{Proceedings of the 7th IFIP Conference on
  Optimization Techniques}} (\bibinfo{publisher}{Springer-Verlag},
  \bibinfo{address}{London, UK}, \bibinfo{year}{1976}), pp.
  \bibinfo{pages}{200--218}.

\bibitem[{\citenamefont{Kaczmarz}(1937)}]{Kaczmarz1937}
\bibinfo{author}{\bibfnamefont{S.}~\bibnamefont{Kaczmarz}},
  \bibinfo{journal}{Bull. Int. Acad. Polon. Sci. A}
  \textbf{\bibinfo{volume}{35}}, \bibinfo{pages}{355} (\bibinfo{year}{1937}).

\bibitem[{\citenamefont{Elser}(2003)}]{elser2003pri}
\bibinfo{author}{\bibfnamefont{V.}~\bibnamefont{Elser}}, \bibinfo{journal}{J.
  Opt. Soc. Am.} \textbf{\bibinfo{volume}{20}}, \bibinfo{pages}{40}
  (\bibinfo{year}{2003}).

\bibitem[{\citenamefont{Thibault}(2007)}]{ThesePT}
\bibinfo{author}{\bibfnamefont{P.}~\bibnamefont{Thibault}}, Ph.D. thesis,
  \bibinfo{school}{Cornell University} (\bibinfo{year}{2007}).

\bibitem[{\citenamefont{Elser and Rankenburg}(2006)}]{ivan1}
\bibinfo{author}{\bibfnamefont{V.}~\bibnamefont{Elser}} \bibnamefont{and}
  \bibinfo{author}{\bibfnamefont{I.}~\bibnamefont{Rankenburg}},
  \bibinfo{journal}{Phys. Rev. E} \textbf{\bibinfo{volume}{73}},
  \bibinfo{pages}{26702} (\bibinfo{year}{2006}).

\bibitem[{\citenamefont{Rankenburg and Elser}(2007)}]{ivan2}
\bibinfo{author}{\bibfnamefont{I.}~\bibnamefont{Rankenburg}} \bibnamefont{and}
  \bibinfo{author}{\bibfnamefont{V.}~\bibnamefont{Elser}},
  \bibinfo{journal}{arXiv: 0706.1754}  (\bibinfo{year}{2007}).

\bibitem[{\citenamefont{Elser et~al.}(2007)\citenamefont{Elser, Rankenburg, and
  Thibault}}]{SSelser2007}
\bibinfo{author}{\bibfnamefont{V.}~\bibnamefont{Elser}},
  \bibinfo{author}{\bibfnamefont{I.}~\bibnamefont{Rankenburg}},
  \bibnamefont{and} \bibinfo{author}{\bibfnamefont{P.}~\bibnamefont{Thibault}},
  \bibinfo{journal}{PNAS} \textbf{\bibinfo{volume}{104}}, \bibinfo{pages}{418}
  (\bibinfo{year}{2007}).

\bibitem[{\citenamefont{Cheeseman et~al.}(1991)\citenamefont{Cheeseman,
  Kanefsky, and Taylor}}]{cheeseman1991rhp}
\bibinfo{author}{\bibfnamefont{P.}~\bibnamefont{Cheeseman}},
  \bibinfo{author}{\bibfnamefont{B.}~\bibnamefont{Kanefsky}}, \bibnamefont{and}
  \bibinfo{author}{\bibfnamefont{W.}~\bibnamefont{Taylor}},
  \bibinfo{journal}{Proceedings of the 12th IJCAI} pp.
  \bibinfo{pages}{331--337} (\bibinfo{year}{1991}).

\bibitem[{\citenamefont{Conway and Sloane}(1999)}]{conway1999}
\bibinfo{author}{\bibfnamefont{J.}~\bibnamefont{Conway}} \bibnamefont{and}
  \bibinfo{author}{\bibfnamefont{N.}~\bibnamefont{Sloane}},
  \emph{\bibinfo{title}{{Sphere Packings, Lattices and Groups}}}
  (\bibinfo{publisher}{Springer}, \bibinfo{year}{1999}).

\bibitem[{\citenamefont{Stephenson}(2005)}]{stephenson2005icp}
\bibinfo{author}{\bibfnamefont{K.}~\bibnamefont{Stephenson}},
  \emph{\bibinfo{title}{{Introduction to Circle Packing}}}
  (\bibinfo{publisher}{Cambridge University Press}, \bibinfo{year}{2005}).

\bibitem[{\citenamefont{Szab{\'o} et~al.}(2007)\citenamefont{Szab{\'o},
  Mark{\'o}t, Csendes, Specht, Casado, and Garc{\~a}a}}]{szabo2007}
\bibinfo{author}{\bibfnamefont{P.}~\bibnamefont{Szab{\'o}}},
  \bibinfo{author}{\bibfnamefont{M.}~\bibnamefont{Mark{\'o}t}},
  \bibinfo{author}{\bibfnamefont{T.}~\bibnamefont{Csendes}},
  \bibinfo{author}{\bibfnamefont{E.}~\bibnamefont{Specht}},
  \bibinfo{author}{\bibfnamefont{L.}~\bibnamefont{Casado}}, \bibnamefont{and}
  \bibinfo{author}{\bibfnamefont{I.}~\bibnamefont{Garc{\~a}a}},
  \emph{\bibinfo{title}{{New Approaches to Circle Packing in a Square}}}
  (\bibinfo{publisher}{Springer-Verlag, New York}, \bibinfo{year}{2007}).

\bibitem[{\citenamefont{Nurmela}(1995)}]{nurmela1995csc}
\bibinfo{author}{\bibfnamefont{K.}~\bibnamefont{Nurmela}},
  \emph{\bibinfo{title}{{Constructing Spherical Codes by Global Optimization
  Methods}}} (\bibinfo{publisher}{Helsinki University of Technology},
  \bibinfo{year}{1995}).

\end{thebibliography}

\end{document}